# Structural and electrical properties of ceramic Li-ion conductors based on $Li_{1.3}Al_{0.3}Ti_{1.7}(PO_4)_3$-LiF

K. Kwatek[1*], W. Ślubowska[1], J. Trébosc[2,3], O. Lafon[2,4], J.L. Nowiński[1]

[1] Warsaw University of Technology, Faculty of Physics, 00-662 Warsaw, Poland

[2] Univ. Lille, CNRS, Centrale Lille, ENSCL, Univ. Artois, UMR 8181 – UCCS – Unité de Catalyse et Chimie du Solide, F-59000 Lille, France

[3] Univ. Lille, CNRS-FR2638, Fédération Chevreul, F-59000 Lille, France

[4] Institut Universitaire de France, 1 rue Descartes, F-75231 Paris Cedex 05, France

*Corresponding author. E-mail address: konrad.kwatek@pw.edu.pl

**Abstract**

The work presents the investigations of $Li_{1.3}Al_{0.3}Ti_{1.7}(PO_4)_3$-$x$LiF Li-ion conducting ceramics with $0 \leq x \leq 0.3$ by means of X-ray diffractometry (XRD), $^7$Li, $^{19}$F, $^{27}$Al and $^{31}$P Magic Angle Spinning Nuclear Magnetic Resonance (MAS NMR) spectroscopy, thermogravimetry (TG), scanning electron microscopy (SEM), impedance spectroscopy (IS) and density method. It has been shown that the total ionic conductivity of both as-prepared and ceramic $Li_{1.3}Al_{0.3}Ti_{1.7}(PO_4)_3$ is low due to a grain boundary phase exhibiting high electrical resistance. This phase consists mainly of berlinite crystalline phase as well as some amorphous phase containing $Al^{3+}$ ions. The electrically resistant phases of the grain boundary decompose during sintering with LiF additive. The processes leading to microstructure changes and their effect on the ionic properties of the materials are discussed in the frame of the brick layer model (BLM). The highest total ionic conductivity at room temperature was measured for LATP-0.1LiF ceramic sintered at 800°C and was equal to $\sigma_{tot} = 1.1 \times 10^{-4}$ S·cm$^{-1}$.

## 1. Introduction

Solid lithium ion electrolytes are interesting materials for the Li-ion battery technology (LIB). When applied, they offer not only improved mechanical and thermal properties of the batteries but also possibility to use new high-voltage (> 5 V) cathode materials in these devices [1-4]. The principal criterion to classify a lithium ion conductor as applicable for the LIB's technology is the value of the total ionic conductivity at room temperature, which should exceed $10^{-4}$ S·cm$^{-1}$ [1-4]. Unfortunately, most of the materials, which fulfills this conductivity condition suffers from poor chemical stability in the presence of moisture. The glasses based on Li$_2$S suffer from such limitation [2,4]. On the other hand, the materials demonstrating a high chemical stability, e.g. lithium ion conducting ceramic materials based on metal oxides, exhibit poor total ionic conductivity. Various attempts have been made to obtain a solid electrolyte, which combines both high lithium ion conductivity and high chemical stability against moisture [2,4]. One of the possible approaches consists in modifying the above-mentioned ceramic materials based on metal oxides to improve their ionic conductivity. Their low total ionic conductivity mainly results from the grain boundary with high electric resistance surrounding highly conducting grains. Thus, to obtain a good solid electrolyte, it is necessary to modify the grain boundary in such a way that its resistance is reduced significantly.

Within the ceramic materials based on metal oxides, which are considered as the candidates for the modification of their grain boundaries, lithium titanium phosphate (LTP) with the chemical formula $LiTi_2(PO_4)_3$ seems to be promising [5-12]. The compound crystallizes in a NASICON-type structure with rhombohedral symmetry of R-3c space group. The bulk (grain) conductivity of LTP is in the range of $10^{-6}$-$10^{-3}$ S·cm$^{-1}$ at room temperature [5-12]. The derived compounds, in which some of $Ti^{4+}$ ions in the crystal structure are substituted by $Al^{3+}$, show even higher ion conductivity. Among these compounds, the $Li_{1+x}Al_xTi_{2-x}(PO_4)_3$ with $x = 0.3$, called LATP hereafter, received much attention, as it exhibits the best ionic properties in the NASICON-based family [5-18]. The LATP materials obtained via melt-quenching or sol-gel method exhibit total ionic conductivity up to $10^{-3}$ S·cm$^{-1}$ at room temperature [19-23].

Although the modification of the starting material should obviously be made at the technology stage of the synthesis or sintering, the issue how to do it efficiently is still under discussion. In the literature, it has been reported that the sintering in the presence of an additional foreign phase can improve the ionic properties [17,24-31]. N. C. Rosero-Navarro *et al.* investigated the addition of $Li_3BO_3$ to garnet type materials and showed that this addition increased the total ionic conductivity of the formed composites [24]. They also noticed the densification of the ceramic materials after sintering at 900°C. Furthermore, H. Aono *et al.* observed an improvement of the total ionic conductivity of the studied $LiTi_2(PO_4)_3$-$Li_3BO_3$ and $LiTi_2(PO_4)_3$-$Li_3PO_4$ composites in comparison to the pristine $LiTi_2(PO_4)_3$ material [25]. Besides, in our previous works related to LTP-based composites, to which $Li_3BO_3$ or $Li_{2.9}B_{0.9}S_{0.1}O_{3.1}$ glass have been added, we observed a significant enhancement of the total ionic conductivity by almost four orders of magnitude with respect to ceramic $LiTi_2(PO_4)_3$ [26,27]. T. Hupfer *et al*. studied the LATP based composite with the addition of different amounts of $LiTiOPO_4$ [28]. They reported that the secondary phase lowered the sintering temperatures. Moreover, the assistance of the foreign phases caused densification of the studied materials. It also affected the total ionic conductivity of the composite. N. Kyono et al. [17] presented that even the addition of an insulating phase in an appropriate amount followed by sintering improved the total ionic conductivity of NASICON-family material.

Among various sintering additives, lithium fluoride appears to be an interesting and effective one [29-31]. B. Xu *et al.* investigated the perovskite-type ceramics with the chemical formula $Li_{3/8}Sr_{7/16}Hf_{1/4}Ta_{3/4}O_3$ (LSHTO) [29]. The authors observed that LiF additive increased the total ionic conductivity of LSHTO by about 1.5 times with respect to the starting material, yielding a conductivity of $3.3 \times 10^{-4}$ S·cm$^{-1}$ at room temperature. L. Xiong *et al.* reported another approach to enhance the ionic properties of lithium titanium phosphate using lithium fluoride additive [30]. Instead of sintering the as-prepared material with LiF, the additive was added to substrates before the synthesis process. They observed a significant enhancement of total ionic conductivity, which was ascribed to the changes in the microstructure, i.e. bigger and better compacted grains. In ref. [31], we described the LTP-LiF composite family and measured a total ionic conductivity enhanced by three orders of magnitude with respect to LTP.

Although all the literature reports unanimously emphasize the importance of LiF additive to increase total ionic conductivity, detailed explanations are still lacking. This work aims at better understanding the role of LiF additive and sintering temperature in the improvement of the total ionic conductivity of the $Li_{1.3}Al_{0.3}Ti_{1.7}(PO_4)_3$-LiF composites. For this purpose, we decided to investigate not only LATP-LiF ceramic composites sintered at 700, 800 and 900°C but also the as-received LATP powder and LATP ceramics after sintering at 700, 800 and 900°C, which were considered as reference materials. These materials were investigated using: high temperature X-ray diffractometry (HTXRD), scanning electron microscopy (SEM), $^7$Li, $^{19}$F, $^{27}$Al and $^{31}$P magic angle spinning nuclear magnetic resonance spectroscopy (MAS NMR), impedance spectroscopy (IS), thermal gravimetric analysis (TGA) and density method. We paid attention to correlations among the results obtained by the means of the methods used for characterization of the studied materials.

## 2. Experimental details

The polycrystalline $Li_{1.3}Al_{0.3}Ti_{1.7}(PO_4)_3$ was obtained by the means of solid-state reaction method. Starting reagents $Li_2CO_3$ (Sigma Aldrich), $NH_4H_2PO_4$ (POCh), $TiO_2$ (Sigma Aldrich) and $Al_2O_3$ (Sigma Aldrich) were weighted in stoichiometric amounts, then mixed with a mortar and pestle. The ground mixture was annealed at 900°C for 10 h to obtain the final compound. In the next stage of preparation of composites, a ball-milling technique was employed for mixing the as-prepared LATP powder with LiF taken in molar ratio varying from 10 to 30 mol%. Both powders, immersed in ethanol, were ball-milled with rotation speed of 400 rpm for 1 h using a planetary ball-mill Fritsch Pulverisette 7. The obtained powder, after drying, was pressed under uniaxial 10 MPa pressure to form pellets 6 mm in diameter and ca. 2 mm thick. Finally, they were sintered at 700, 800 or 900°C for 2 h.

The X-ray diffraction method employing Philips X'Pert Pro (Cu Kα) was used for examination of the quality of both the as-synthesized and composite powdered products. Temperature dependent XRD (HTXRD) was performed using Anton Paar HTK-1200 Oven in the temperature 30 - 900°C range. Data were collected in the range of 10° - 90° with 0.05° step and a count rate of 0.5 s at each step.

Thermal gravimetric analysis (TGA) was carried out on the powdered composites using TA Instruments Q600 to observe mass loss during heating under air flow in temperature range of 30 to 900°C. The measurements were performed with a heating rate of 10°C.min$^{-1}$ on samples of ca. 20 mg each. The investigation aimed at determining the thermal stability of the composites.

The apparent density of the composites was determined using Archimedes method with isobutanol as an immersion liquid. We estimated the accuracy of the used method as ca. 1%. The microstructure was investigated by means of SEM employing Raith eLINE plus. For the SEM imaging, always freshly fractured pellets were used.

For electric measurements, both bases of the as-formed pellets were, at first, polished and then covered with graphite as electrodes. Impedance investigations were carried out employing Solatron 1260 frequency analyzer in a frequency range of $10^{-1}$-$10^{7}$ Hz. Impedance data were collected in the temperature range from 30 to 100°C, both during heating and cooling runs.

MAS NMR spectra were acquired on a 400 MHz Bruker Avance II NMR spectrometer equipped with a 4 mm HXY probe used in double resonance mode. The NMR spectra were acquired at room temperature (RT) at the MAS frequency of 10 kHz. Single-pulse experiments were performed with pulse lengths of 1.0 μs for $^{7}$Li, 1.0 μs for $^{27}$Al and 2.5 μs for $^{31}$P. Radiofrequency (rf) nutation frequencies were equal to 115 kHz for $^{7}$Li, 60 kHz for $^{27}$Al and 50 kHz for $^{31}$P. The $^{7}$Li, $^{27}$Al and $^{31}$P NMR spectra result from the averaging of 512, 128, 8 transients using relaxation delays of 0.3, 1.0 or 20 s. The $^{19}$F NMR spectra were acquired at room temperature and a MAS frequency of 30 kHz using single-pulse experiments with a pulse length of 2.8 μs and an rf nutation frequency of 89 kHz. The $^{19}$F NMR spectrum resulted from averaging 144 transients with a relaxation delay of 10 s. The $^{7}$Li, $^{27}$Al and $^{31}$P chemical shifts were referenced to 1 mol.L$^{-1}$ LiCl, 1 mol.L$^{-1}$ AlCl$_3$ and 85 wt% H$_3$PO$_4$ aqueous solutions. The $^{19}$F chemical shifts were referenced to CFCl$_3$ using the resonance of CaF$_2$ (−108.6 ppm) as secondary reference. The NMR spectra were simulated using dmfit software [32]. The $^{31}$P and $^{27}$Al NMR spectra were simulated assuming Gaussian lineshapes and considering only the isotropic shifts, whereas the $^{7}$Li NMR spectra were simulated assuming Lorentzian lineshapes for all bands and considering isotropic chemical shift and quadrupolar interaction.

## 3. Results and discussion

### *3.1 X-ray diffraction (XRD)*

Fig. 1 presents the X-ray diffraction patterns for the as-prepared, polycrystalline LATP as well as the LATP-0.3LiF composite before and after sintering at 900°C. The position and relative intensity of the main XRD reflections for the as-prepared LATP powder, correspond to the NASICON-type compounds with R-3c symmetry group. Besides them, some additional weak diffraction lines located at 2θ angles 20.6° and 26.3° can be observed. They were ascribed to the berlinite aluminophosphate phase (denoted $AlPO_4$). In the case of the non-sintered composite, the XRD pattern is similar and the only additional reflections being weak peaks at 38.6° and 45.0° assigned to the LiF additive. After sintering at 900°C, some new weak reflections are observed at the following 2θ angles: 22.2, 26.7, 27.7, 31.5 and 39.3°. They were assigned to $LiTiPO_5$ and $Li_4P_2O_7$ phases. Furthermore, the diffraction peaks attributed to the LiF phase disappeared. To determine the temperatures at which the above phase transformations occur, the HTXRD was performed. During annealing, above 300°C, the intensity of the LiF diffraction peaks started to decrease with increasing temperature and finally at about 500°C the peaks completely vanished. Next changes were observed when temperature reached 700°C. The diffraction peaks related to $AlPO_4$ phase faded out and simultaneously those attributed to the $LiTiPO_5$ and $Li_4P_2O_7$ appeared. The results suggested that at high temperatures the $AlPO_4$ phase reacted with the LATP and resulted in the formation of new compounds, including $LiTiPO_5$ and $Li_4P_2O_7$ crystalline phases.

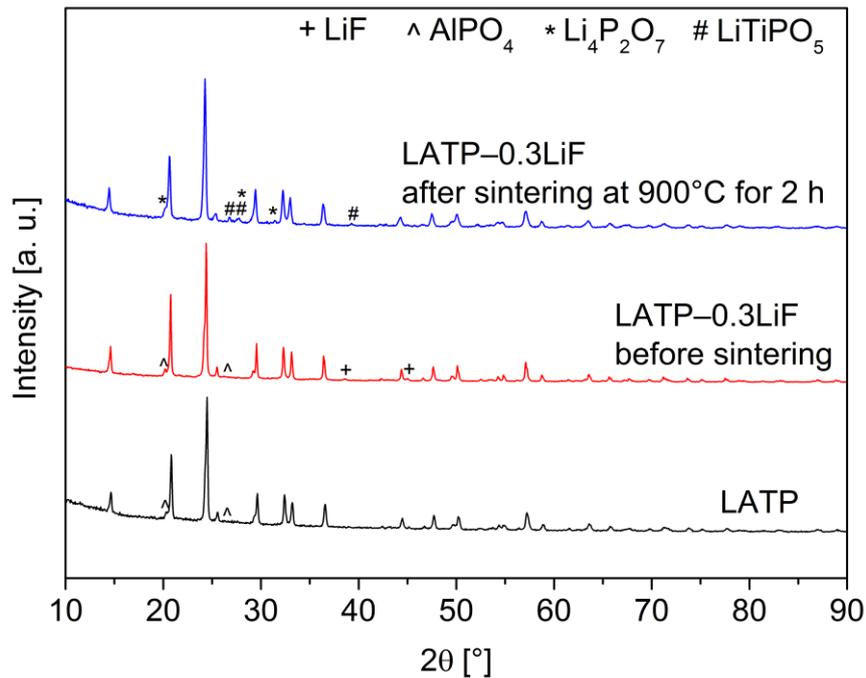

Figure 1 XRD patterns of the as-prepared LATP as well as the LATP-0.3LiF composite before and after sintering at 900°C.

## 3.2 Thermal gravimetric analysis (TGA)

The thermal gravimetric (TG) results of the studied composites are presented in Fig. 2. Between RT and about 500°C, for all the investigated materials with different LiF molar ratios, mass loss occurs, whereas above 500°C there is no significant mass loss. The mass loss from RT to about 200°C stems from the evaporation of the moisture and residual ethanol adsorbed on the surface of the grains. At higher temperatures, the decomposition of LiF and the associated release of the fluorinated compound produces additional mass loss.

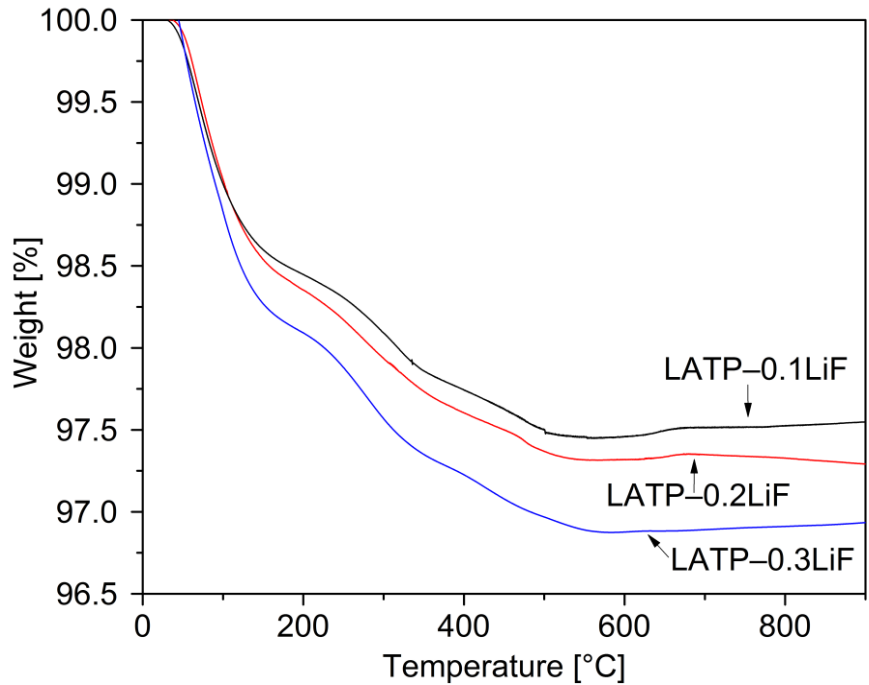

Figure 2 TG traces for LATP composites formed with various molar ratios of LiF.

### 3.3 *SEM and density*

Fig. 3 displays the SEM images of the surface of the freshly fractured pellets of LATP ceramic sintered at 900°C as well as LATP-0.2LiF and sintered at 700, 800 or 900°C. Fig. 3A shows the SEM image of the ceramic LATP material, which is formed of two kinds of grains: small grains with a diameter of ca. 1.5 μm as well as bigger ones with a size of a few micrometers. In addition, some voids, microcracks and grain boundaries are visible. The microstructure of the composite containing 0.2 mol of LiF and sintered at 700°C is different (Fig. 3B). First of all, this sample is made mainly of grains with a size of ca. 1.5 μm. Additionally, the concentration of pores is higher than in the pristine material and above all, the LATP grains seem to adhere better to each other. When the composite is sintered at 800°C, the grains become bigger and more densely packed (Fig. 3C). No large voids are observed. After sintering at 900°C (Fig. 3D), the grain size exceeds 5 μm. They are well matched to each other, however some microcracks are observed. In summary, the SEM investigations show that LiF acts as ceramic densification agent. The LATP grains in the ceramic composite are bigger and more densely packed with sintering temperature.

The results of the density measurements for the LATP and studied composites are listed in Table 1. The apparent densities of the composites vary in the narrow range of 2.71-2.79 g·cm$^{-3}$. They are slightly larger than those of the pristine material sintered at the same temperature and slightly increase with sintering temperature. The obtained data shows a negligible influence of LiF on the density values. The relative density values were calculated in regards to theoretical density of LATP (2.946 g·cm$^{-3}$) [11,21].

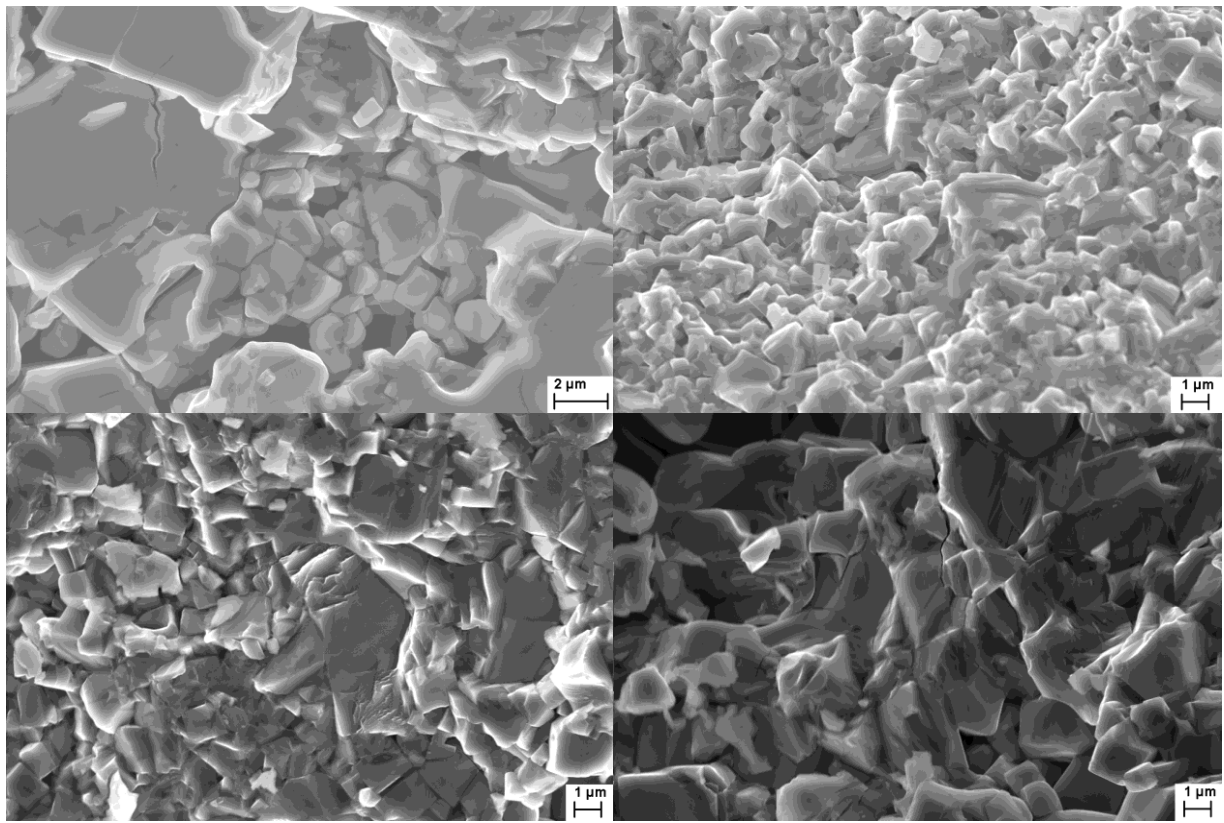

Figure 3 SEM images for (A) the LATP sintered at 900°C, (B-D) the LATP-0.2LiF sintered at (B) 700°C, (C) 800°C and (D) 900°C.

Table 1 Values of the apparent and relative density of the materials sintered at different temperatures.

| Composite | Sintering temperature [°C] | Apparent density [g·cm$^{-3}$] | Relative density [%] |
|---|---|---|---|
| LATP | 700 | 2.69 ± 0.03 | 91.3 ± 1.2 |
| | 800 | 2.70 ± 0.03 | 91.6 ± 1.0 |
| | 900 | 2.71 ± 0.04 | 92.1 ± 1.6 |
| LATP-0.1LiF | 700 | 2.70 ± 0.03 | 91.8 ± 1.0 |
| | 800 | 2.78 ± 0.03 | 94.2 ± 0.9 |
| | 900 | 2.80 ± 0.03 | 95.0 ± 0.9 |
| LATP-0.2LiF | 700 | 2.71 ± 0.04 | 92.0 ± 1.4 |
| | 800 | 2.77 ± 0.03 | 93.9 ± 1.0 |
| | 900 | 2.79 ± 0.03 | 94.6 ± 0.9 |
| LATP-0.3LiF | 700 | 2.71 ± 0.03 | 92.0 ± 1.1 |
| | 800 | 2.76 ± 0.04 | 93.8 ± 1.4 |
| | 900 | 2.79 ± 0.03 | 94.8 ± 1.1 |

### *3.4 MAS NMR*

The MAS NMR investigations were focused on the as-prepared LATP powder, the LATP ceramic sintered at 900°C, the LATP-0.3LiF composite non-sintered and sintered at 800°C.

### *3.4.1 $^{19}$F MAS NMR*

The $^{19}$F MAS NMR spectrum (Fig. 4) of the non-sintered LATP-0.3LiF sample is dominated by a signal with a centerband at −204 ppm assigned to LiF [33]. An additional weak signal with a centerband at −165 ppm is also observed. This peak may result from the formation of Al-F bonds in LATP-0.3LiF [34]. Conversely after sintering, no $^{19}$F NMR signal is detected. These results are consistent with the HTXRD and TG data and indicate that LiF decomposes during sintering with an associated release of fluorine atoms from the material.

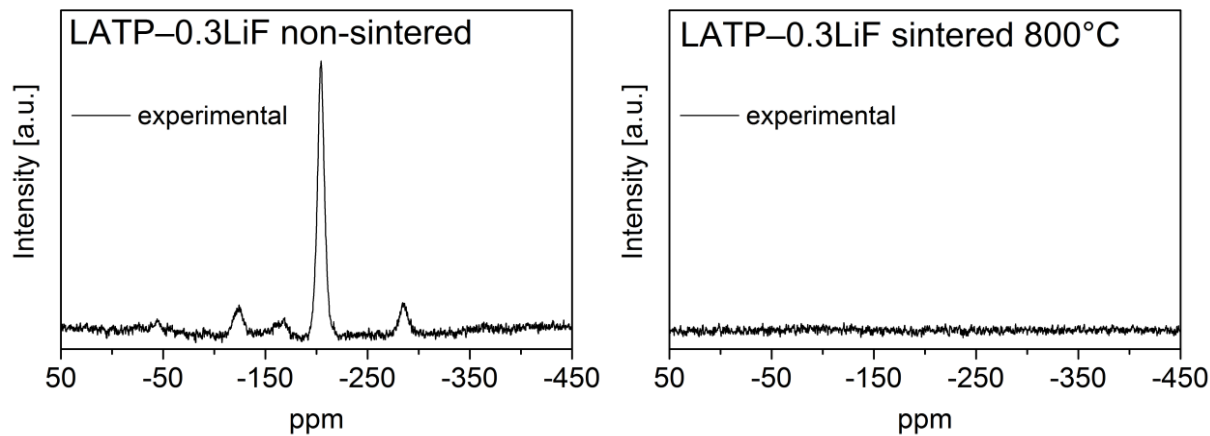

Figure 4 $^{19}$F MAS NMR spectra of LATP-LiF non-sintered and sintered at 800°C. The experimental spectra are displayed as black lines.

### 3.4.2 $^{27}$Al MAS NMR

The $^{27}$Al MAS NMR spectra (Fig. 5) for the non-sintered as-prepared LATP show three peaks at −15, 13 and 40 ppm assigned to the hexa- (AlO$_6$), penta- (AlO$_5$) and tetra-coordinated (AlO$_4$) aluminium sites, respectively [7,12,18,23]. NASICON structure contains AlO$_6$ sites and therefore the −15 ppm line was assigned to LATP. The 40 and 13 ppm resonances were assigned to some amorphous or strongly disordered phase containing aluminum, which was detected by the XRD [12,23]. The MAS NMR spectrum of the LATP after sintering at 900°C shows the same lines. However, the sintering modifies their relative integrated intensity (see Table S1). The sintering enhances the relative intensity of NASICON, whereas the relative intensity of the signal resonating at 13 ppm is reduced. Conversely the relative integrated intensity of the line at 40 ppm remains practically unchanged. An additional weak peak at 31 ppm is observed. It was assigned to the berlinite [7,12,18,23,35]. The $^{27}$Al MAS NMR spectra for the non-sintered LATP and non-sintered LATP-0.3LiF are similar. The three $^{27}$Al sites exhibit comparable integrated intensities in both samples. The NMR spectrum of LATP-0.3LiF changes substantially after sintering at

800°C. Only the peak of NASICON structure is visible. This observation clearly indicates that the annealed LATP-0.3LiF is practically free from foreign phases containing $Al^{3+}$ ions.

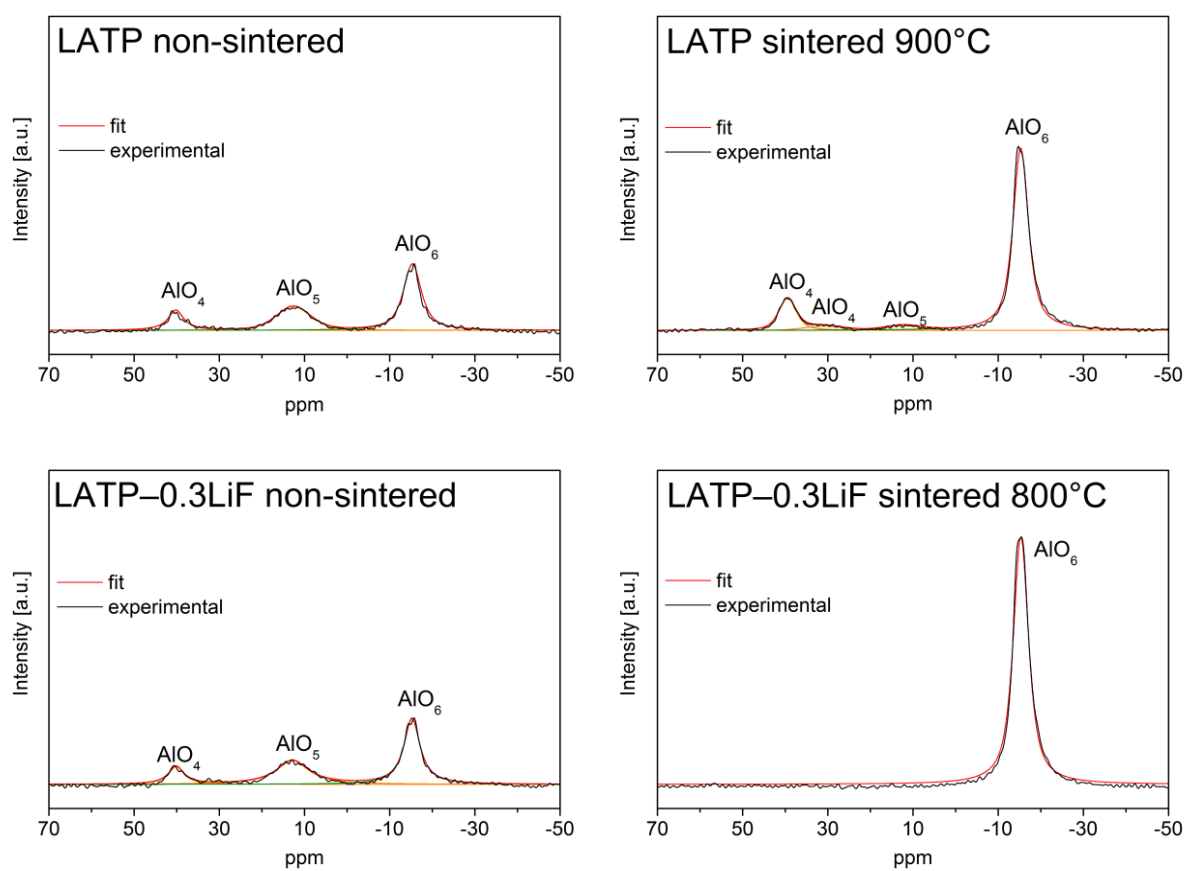

Figure 5 $^{27}$Al MAS NMR spectra of LATP non-sintered and sintered at 900°C as well as LATP-03LiF non-sintered and sintered at 800°C. The experimental and simulated spectra are displayed as black and red lines, respectively. The simulated spectra are the sum of distinct lineshapes displayed as orange and green lines. The best fit parameters are given in Table S1. The spectra of the four samples are displayed with identical intensity scales.

In summary, the $^{27}$Al MAS NMR investigation indicate that the as-prepared material contains, besides LATP and berlinite phases, which have been observed by XRD, some amorphous or highly disordered phase or phases containing aluminum ions, which were not detected by XRD. Annealing at high temperature promotes chemical processes causing destruction of impurity phases accompanied with the transfer of aluminum ions to the crystal lattice of LATP. However, the sintering without LiF assistance does not fully destroy the amorphous phase and the berlinite. The total decomposition of all foreign phases containing $Al^{3+}$ is achieved after the sintering of LATP with LiF.

### 3.4.3 $^{31}P$ MAS NMR

Fig. 6 shows the $^{31}P$ MAS NMR spectrum for the as-prepared LATP, which exhibits a broad asymmetric peak at ca. −27.5 ppm. This peak can be deconvolved into four overlapping line shapes centered at −27.7, −27.0, −26.3, −24.7 ppm and another one, very weak at −18.8 ppm. The latter peak was ascribed to some impurities, plausibly unreacted reactants. The line −24.7 ppm was assigned to phosphorous site in berlinite crystalline phase ($AlPO_4$) [35] or to phosphorous atoms with similar environment in amorphous or highly disordered aluminophosphate phases. The three peaks at −27.7, −27.0 and −26.3 ppm are assigned to the $P(OTi)_4$, $P(OTi)_3(OAl)_1$ and $P(OTi)_2(OAl)_2$ sites, respectively [5,7,10]. For the sintered LATP, the deconvolution of the spectrum indicates the presence of an additional peak at −25.7 ppm, which is assigned to the $P(OTi)_1(OAl)_3$ coordination. The comparison of the relative integrated intensities of the $P(OTi)_{4-x}(OAl)_x$ resonances with $x = 0, 1, 2$ or $3$ between the as-prepared and sintered LATP samples shows the decrease of the $P(OTi)_4$ peak accompanied by the appearance of the $P(OTi)_1(OAl)_3$ peak.

The $^{31}P$ MAS NMR spectra of the non-sintered LATP and LATP-0.3LiF are similar. However, the spectra of the sintered LATP and LATP-0.3LiF differ. No $^{31}P$ signal at −24.7 ppm produced by berlinite or amorphous aluminophoshate phases is visible for the sintered LATP-0.3LiF. This observation agrees with XRD data and the $^{27}Al$ MAS NMR results and confirms decomposition of the berlinite phase and amorphous aluminophosphate phase as the result of sintering. We also noticed a significant decrease of the integrated relative intensity of the $P(OTi)_4$ peak accompanied with an increase of $P(OTi)_1(OAl)_3$ signal and the appearance of P(OAl)$_4$ peak. This conversion is more efficient than for the sintering of LATP. Furthermore, after the sintering, additional peaks at −3.6, −5.9, −9.8 and −23.6 ppm are observed (Fig. 6). The narrow peaks at −3.6, −5.9 and −9.8 ppm are ascribed to $^{31}$P nuclei in some crystalline phosphates, including LiTiPO$_5$ or/and Li$_4$P$_2$O$_7$ phases, which were observed by XRD [36]. The broad peak at −23.6 ppm is ascribed to P atoms linked to three titanium atoms via oxygen bridge in amorphous titanium phosphate (TiPO$_4$) [37]. These phases probably result from decomposition of the aluminophosphate phases. The released aluminium ions diffuse into the bulk of grains, where they substitute titanium ions in the crystal structure. In turn, the released titanium ions diffuse outside the grain and react with PO$_4$ groups and lithium ions from decomposed LiF. The resulting phases deposit on the grain surface.

In the summary, in the as-prepared LATP, only P(OTi)$_{4-x}$(OAl)$_x$ ($x$ = 0, 1 and 2) coordinations are observed and the P(OTi)$_4$ signal dominates. After sintering of both LATP or LATP-LiF, the relative amount of the P(OTi)$_4$ site decreases, while the P(OTi)$_1$(OAl)$_3$ sites are formed. Additionally, in the case of sintered LATP-LiF material, P(OAl)$_4$ coordination is detected. The relative amounts of P(OTi)$_{4-x}$(OAl)$_x$ ($x$ = 1 and 2) remain unchanged during the sintering (see Table S2). As a result, the amount of Al$^{3+}$ ions in the sintered LATP is higher than that in the as-prepared one. Furthermore, the concentration of Al$^{3+}$ ions in the LATP phase is the highest after sintering in the presence of LiF. The result is consistent with the $^{27}$Al MAS NMR observations that indicate the decomposition of the foreign phases containing aluminum ions and the diffusion of the released Al$^{3+}$ ions into the crystal structure of LATP, where they replace Ti$^{4+}$ ions in the crystal lattice.

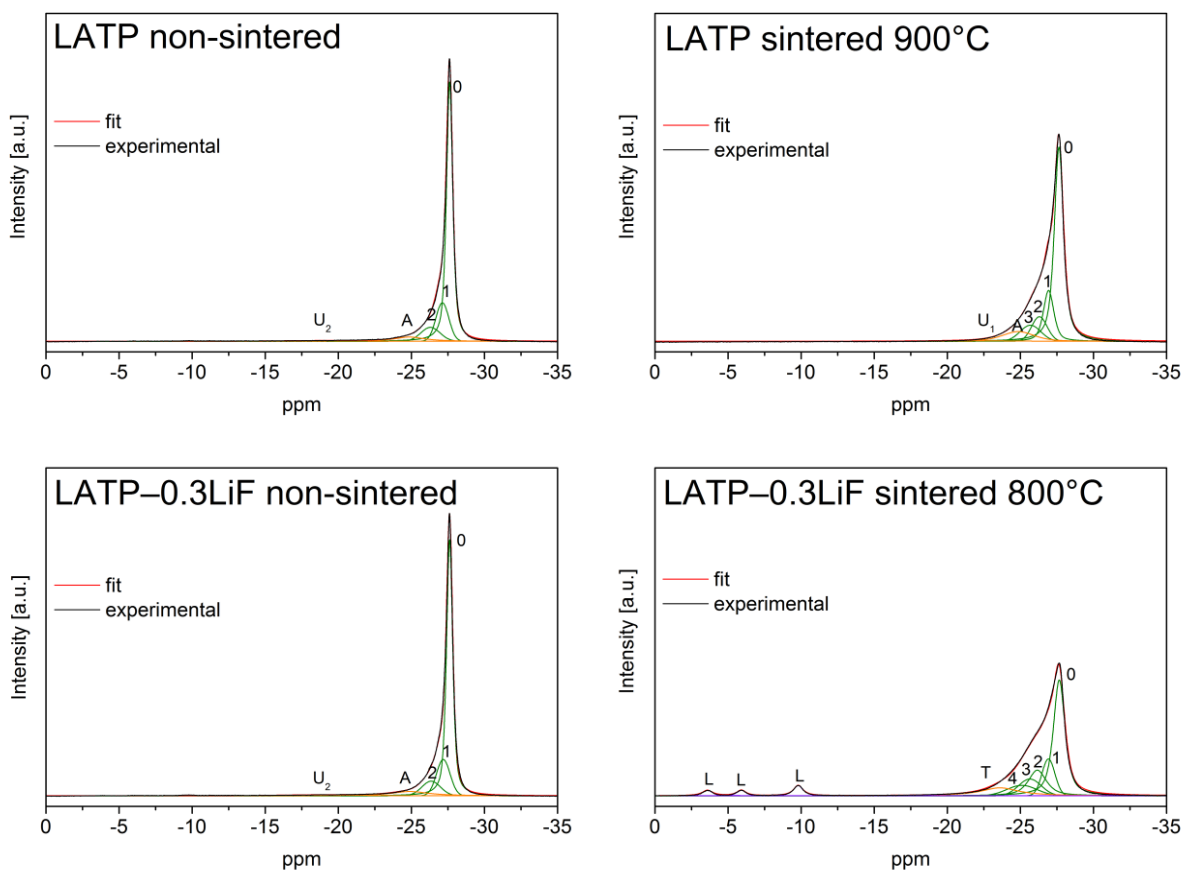

Figure 6 $^{31}$P MAS NMR spectra for LATP non-sintered and sintered at 900°C as well as LATP-03LiF non-sintered and sintered at 800°C. The experimental and simulated spectra are displayed as black and red lines, respectively. The simulated spectra is the sum of distinct lineshapes displayed as orange, green and violet lines. Numbers x = 0-4 stands for P(OTi)$_{4-x}$(OAl)$_x$ sites, A for AlPO$_4$, T for TiPO$_4$, U$_1$ and U$_2$ for impurity phases and L for lithium conducting phosphates. The spectra of the four samples are displayed with identical intensity scales.

### 3.4.4 $^7$Li MAS NMR

Fig. 7 shows the $^7$Li MAS NMR spectrum of the as-prepared LATP. The simulation of the spectrum indicates the presence of two peaks at −1.1 and −0.8 ppm, with quadrupolar coupling constant ($C_Q$) equal to 38 and 71 kHz. These peaks at −1.1 and −0.8 ppm were assigned to the two different lithium sites in the NASICON crystal structure, namely Li1 and Li3 respectively [7,10,18]. The integrated intensities of both peaks are almost equal, indicating equal occupancies of both sites (see Table S3). Nevertheless, after sintering, the $C_Q$ value of both $^7$Li sites decreases. Such decrease may be related to the replacement of Ti$^{4+}$ ions by Al$^{3+}$ ones during sintering. Furthermore, sintering broadens the $^7$Li NMR signals of both sites. This broadening is consistent with an accelerated $^7$Li relaxation after the sintering. Furthermore, the integrated intensity of Li3 site increases, which shows a preferential occupation of the Li3 site. The sintering of LATP-0.3LiF composite produce a similar modification of the $^7$Li spectrum. Hence, the presence of the LiF additive does not affect the mobility of Li$^+$ ions and the occupation rates of the Li sites in LATP.

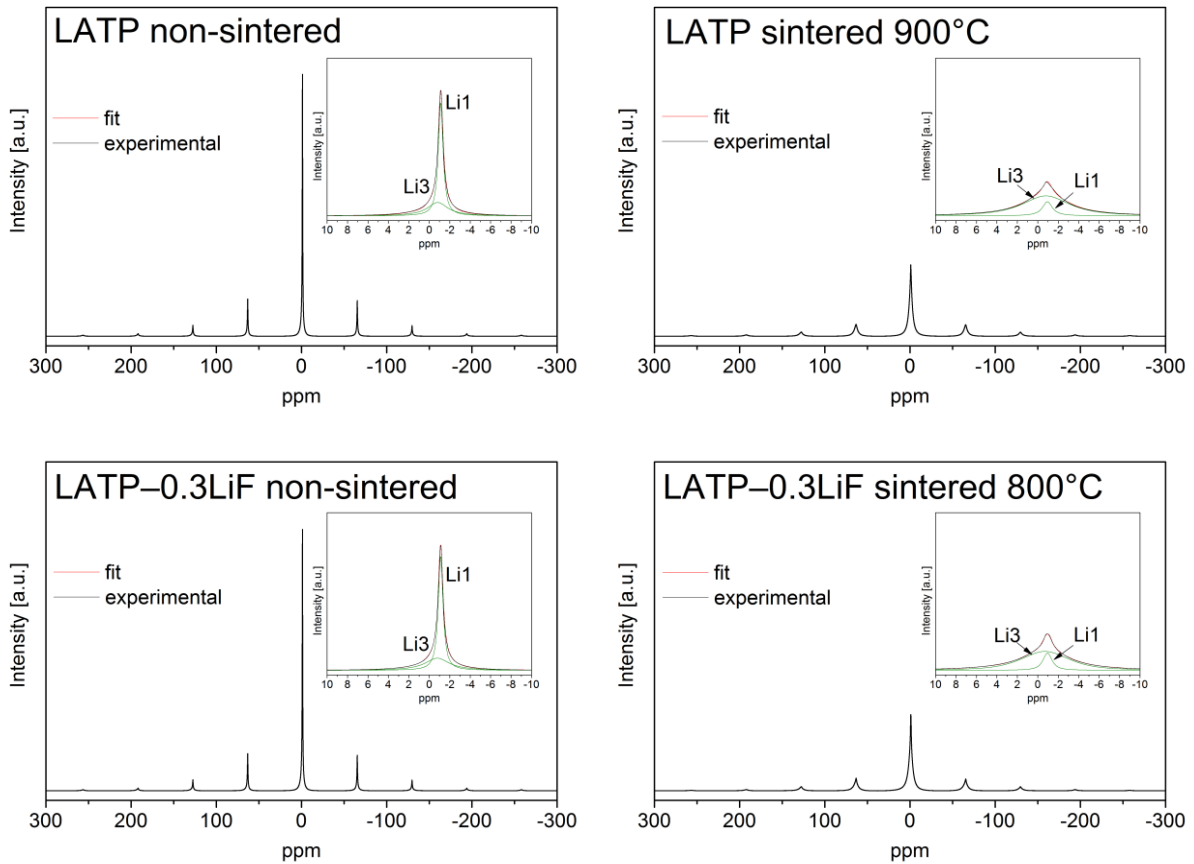

Figure 7 CT of $^7$Li NMR spectra under MAS conditions of LATP non-sintered and sintered at 900°C as well as LATP-03LiF non-sintered and sintered at 800°C. The experimental and simulated spectra are displayed as black and red lines, respectively. The simulated spectra are the sum of distinct lineshapes displayed as orange and green lines. The different sites of LATP are labeled Li1 and Li3. The spectra of the four samples are displayed with identical intensity scales.

### 3.5 Impedance spectroscopy

The results of the impedance spectroscopy investigations for LATP sintered at 900°C and LATP-0.2LiF annealed at 800°C are shown in Fig. 8 in the Nyquist plot representation. The data have been collected for the materials kept at 30°C. The geometries of the samples used in this investigation were similar. The selected impedance plots exhibit typical and characteristic shapes for the studied materials. The Nyquist plot for the ceramic LATP forms a single, almost regular, large semicircle, followed by a spur. Besides that, another, small semicircle can be observed at high frequencies. The plot for the LATP-0.2LiF is representative for the sintered LATP-LiF composite family. It also consists of two semicircles however, the low frequency one is much smaller than that of the LATP ceramic.

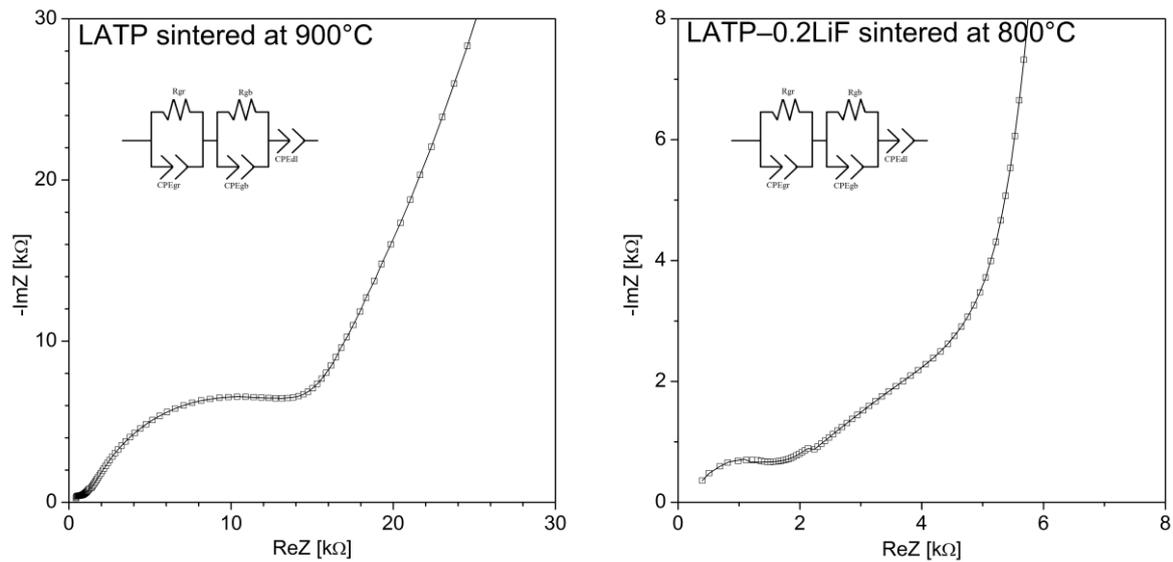

Figure 8 Nyquist plots for the data collected at 30°C for (A) LATP ceramic sintered at 900°C and (B) LATP-0.2LiF composite sintered at 800°C with fitting curves (solid lines) and the equivalent circuit modeling electrical properties of the sintered materials.

The electrical properties of the LATP and derived composites could be modeled via an electrical equivalent circuit approach. An impedance plot, that consists of two separate semicircles is typical for the materials in which ion transport occurs in two different media. In the case of the studied materials, the ions move through the bulk (grains) and inter grain phase (grain boundary). Thus, the corresponding equivalent circuit consists of two resistors: $R_{gr}$ and $R_{gb}$, representing the bulk and grain boundary resistances, respectively. Each resistor is shunted by a constant phase element (CPE). The two resistors are connected in series in the final equivalent circuit. The values of the resistors can be determined from the impedance plot. The intersection point of the low frequency semicircle with Re Z axis determines the total resistance $R_{tot} = R_{gr} + R_{gb}$ value, whereas a similar intersection, but taken for the high frequency semicircle, gives $R_{gr}$. The knowledge of the geometry of the studied samples allows determination of the values of apparent conductivity, which we refer to as conductivity hereafter. The apparent grain, grain boundary and total conductivity values were obtained using the following formula $\sigma = L/(R \cdot A)$, where $L$ and $A$ stand for a thickness and electrode area of the sample respectively. The determined value of the total ionic conductivity of the ceramic LATP sintered at 900°C is ca. $4.7 \times 10^{-5}$ S·cm$^{-1}$ at 30°C. Note that such value is too low for the LIB applications. As seen in Table 2, for materials containing 0.1 or 0.2 mol of LiF and sintered at 800°C, the measured total ion conductivity can be enhanced up to $1.1 \times 10^{-4}$ S·cm$^{-1}$. Temperature dependent impedance spectroscopy shows that for all materials under study, the total conductivity exhibits an Arrhenius dependence (Fig. 9). The estimated values of the activation energy of the total conductivity for various sintering temperatures and LiF contents are reported in the Table 2.

As seen in Table 2, the composites sintered at 800°C exhibit the highest total ion conductivities. The $\sigma_{tot}$ value reaches $1.1 \times 10^{-4}$ S·cm$^{-1}$ for LATP-0.1LiF and LATP-0.2LiF, and is $0.7 \times 10^{-4}$ S·cm$^{-1}$ for LATP-0.3LiF. Furthermore, the materials sintered at 800°C exhibit the lowest values of the $E_{tot}$ equal to 0.23-0.24 eV. This value is significantly lower than those reported for LATP so far (0.38 [11], 0.36 [10] and 0.3 [14] eV). The bulk conductivities, denoted $\sigma_{gr}$ of the composites, vary in a narrow range $(1-3) \times 10^{-4}$ S·cm$^{-1}$. The materials sintered at 800°C or 900°C exhibit identical activation energies of bulk conductivity $E_{gr}$ equal to 0.21 eV, a value which is smaller than those of the composites sintered at 700°C. This value is also smaller than those reported for LATP grains so far (0.29 [11], 0.28 [10] and

0.22 [14]). The conductivity of grain boundary σ$_{gb}$ also vary depending on the sintering temperature and they are also highest for the materials sintered at 800°C.

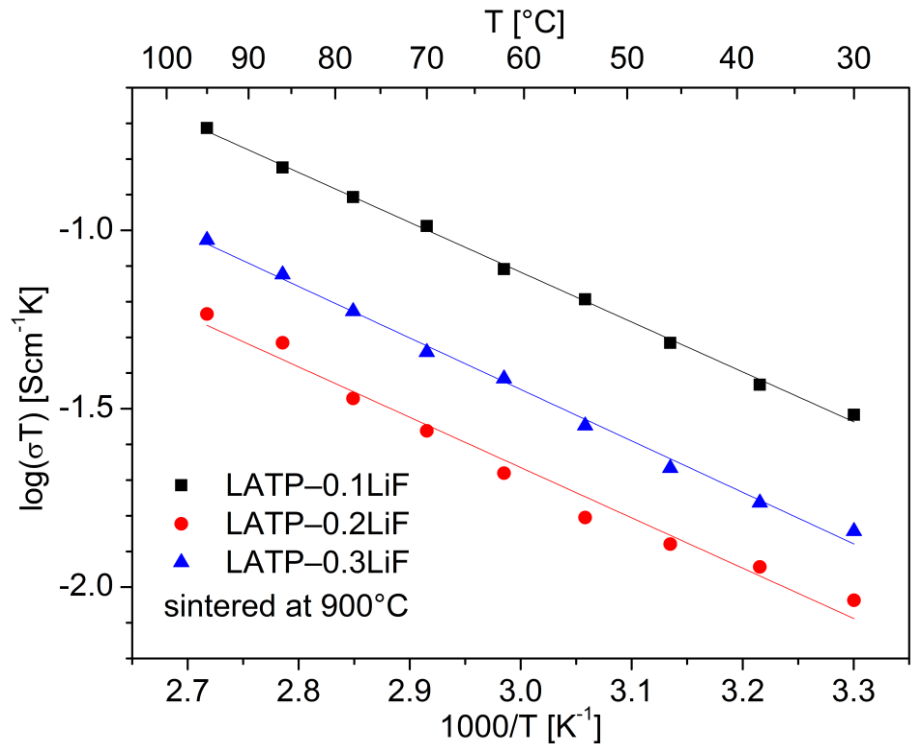

Figure 9 Arrhenius plots of the total electric conductivity of the LATP-LiF composites sintered at 900°C.

Table 2 Values of bulk, grain boundary and total ionic conductivities at 30°C as well as the bulk and total activation energies.

| Composite | $T_{sint}$ [°C] | $\sigma_{gr}$ (30°C) [S·cm$^{-1}$] | $\sigma_{gb}$ (30°C) [S·cm$^{-1}$] | $\sigma_{tot}$ (30°C) [S·cm$^{-1}$] | $E_{gr}$ [eV] | $E_{tot}$ [eV] |
|---|---|---|---|---|---|---|
| LATP | 700 | 8.8 × 10$^{-5}$ | 2.0 × 10$^{-6}$ | 1.9 × 10$^{-6}$ | 0.21 | 0.46 |
| | 800 | 2.0 × 10$^{-4}$ | 1.1 × 10$^{-5}$ | 1.0 × 10$^{-5}$ | 0.21 | 0.39 |
| | 900 | 5.1 × 10$^{-4}$ | 5.2 × 10$^{-5}$ | 4.7 × 10$^{-5}$ | 0.21 | 0.4 |
| LATP-0.1LiF | 700 | 1.1 × 10$^{-4}$ | 1.2 × 10$^{-5}$ | 1.1 × 10$^{-5}$ | 0.24 | 0.27 |
| | 800 | 2.8 × 10$^{-4}$ | 1.8 × 10$^{-4}$ | 1.1 × 10$^{-4}$ | 0.21 | 0.23 |
| | 900 | 3.0 × 10$^{-4}$ | 1.5 × 10$^{-4}$ | 1.0 × 10$^{-4}$ | 0.21 | 0.28 |
| LATP-0.2LiF | 700 | 1.3 × 10$^{-4}$ | 6.1 × 10$^{-5}$ | 4.2 × 10$^{-5}$ | 0.25 | 0.28 |
| | 800 | 2.8 × 10$^{-4}$ | 1.7 × 10$^{-4}$ | 1.1 × 10$^{-4}$ | 0.21 | 0.24 |
| | 900 | 2.0 × 10$^{-4}$ | 3.6 × 10$^{-5}$ | 3.0 × 10$^{-5}$ | 0.21 | 0.29 |
| LATP-0.3LiF | 700 | 2.3 × 10$^{-4}$ | 9.3 × 10$^{-5}$ | 6.6 × 10$^{-5}$ | 0.24 | 0.28 |
| | 800 | 2.0 × 10$^{-4}$ | 1.1 × 10$^{-4}$ | 7.0 × 10$^{-5}$ | 0.21 | 0.23 |
| | 900 | 1.6 × 10$^{-4}$ | 6.7 × 10$^{-5}$ | 4.7 × 10$^{-5}$ | 0.21 | 0.28 |

More detailed analysis of the ionic properties of the composites were performed using the brick-layer model (BLM) [38-43]. The irregular morphology of the microstructure of the real ceramic is modeled as identical regularly arranged cubes representing the grains. Each cube with an edge $D$ is coated by a thin layer with a thickness $d$, which models the grain boundary phase. The total conductivity for such system can be expressed according to the following analytical formula [38]:

$$\sigma_{tot} = \frac{(1-2\alpha)^2 \sigma_{gr}}{(1-2\alpha)+2\alpha\frac{\sigma_{gr}}{\sigma_{gb}}} + 4\alpha(1-\alpha)\sigma_{gb} \qquad (1)$$

where, $\alpha = d/D$. In the above equation, $\sigma_{gr}$ and $\sigma_{gb}$ denote to the true conductivities of grains and grain boundary phases respectively, but not the apparent ones determined by the means of impedance spectroscopy. When $\sigma_{gr} \gg \sigma_{gb}$ i.e. conductivity of the grain is much higher than that of grain boundary, the formula (1) can be simplified to:

$$\sigma_{tot} \approx \left[\frac{(1-2\alpha)^2}{2\alpha} + 4\alpha(1-\alpha)\right]\sigma_{gb} \qquad (2)$$

Hence, the total ion conductivity is proportional to $\sigma_{gb}$ and the proportionality factor (in brackets) depends only on the geometry of the cubes modelling the microstructure. Such situation is encountered for the as-prepared and sintered LATP as well as non-sintered LATP-LiF composites, for which $\sigma_{gr} > 10\sigma_{gb}$.

These results are consistent with XRD and NMR data. These techniques demonstrated that the as-prepared LATP and non-sintered composite contain not only LATP but also poorly ion conductive phases, such as berlinite and some amorphous phase containing $Al^{3+}$ ions. These foreign phases embedded in the LATP grains form a highly resistant medium, which is detected and identified by the means of impedance spectroscopy method as a grain boundary. A sintering without LiF addition causes the densification of the material and the decomposition of the amorphous phase but leaves $AlPO_4$ phase unaltered. Therefore, the total ion conductivity of the sintered LATP ceramic remains low, even if it is much higher than that of the as-prepared non-sintered LATP pellets.

A sintering with the LiF additive decomposes all highly resistant phases within the grain boundary, i.e. the amorphous as well as berlinite. At their expense, some other lithium ion conducting foreign phases are formed, but in small amounts. As a result, the real conductivity of the grain boundary significantly increases. Furthermore, sintering with LiF also alters the microstructure of the material. Grains with size in the order of a micrometer are formed after sintering at 800°C. Sintering at 900°C can further increase the size of the grains.

This morphology corresponds to long grain edge with respect to the layer thickness in BLM model, i.e. $\alpha \ll 1$. Under this assumption Eq. 1 can be recast as:

$$\sigma_{tot} = \frac{\sigma_{gr}}{1+2\alpha\frac{\sigma_{gr}}{\sigma_{gb}}} \qquad (3)$$

Hence, the BLM predicts that the total electric conductivity of the ceramic composed of large grains is proportional to the conductivity of a grain. The proportionality factor depends on both the geometry of the microstructure (parameter $\alpha$) and conductivity ratio $\sigma_{gr}/\sigma_{gb}$. It increases when the real conductivity of grain boundary increases. If $\alpha\sigma_{gr}/\sigma_{gb} \ll 1$, i.e. the grains are large and the real conductivities of grains and grain boundary are comparable, the total conductivity of the ceramic approaches the conductivity of the grains.

BLM and Eqs. 1-3 emphasize the importance of the microstructure and the real conductivity of the grain boundary in order to obtain ceramics with high total ion conductivity. Recently, A. Vyalikh *et al.* [44] reported the significant impact of those factors in enhancement of total ionic conductivity of LAGPY material. The brick-layer model also shows the coupling between these two factors. Notably the negative effect of highly resistant grain boundary on the total ion conductivity can be reduced by an enlargement of the grains. This approach is exploited in the present work. However, in the present work, XRD and NMR characterization have also shown that the sintering affected the grains themselves. The observations indicated that the preparation method of the LATP powder, led to incomplete synthesis of the final product. The resultant LATP powder still contained residues of the unreacted starting materials, some amounts of intermediates and by-products. Therefore, the concentration of the aluminum ions in the LATP crystal lattice was lower than expected. The sintering allowed the formation of grains with a chemical composition closer to the assumed one. Various coupled processes occurs during the sintering: the generation of the free aluminum ions on a grain surface after decomposition of the inter-grain phases, the diffusion of the aluminum ions into the bulk, the substitution of $Ti^{4+}$ ions in the crystal lattice by $Al^{3+}$ ones, the backward diffusion of the released titanium ions to the grain surface, a reaction of the titanium ions with the inter-grain material and finally the modification and the formation of new phases as grain boundaries. All these processes promoted grain growth. The temperature affects these processes. At 700°C, the substitution process of $Ti^{4+}$ by aluminum ion and the modification of the grain boundary were not fully completed. Therefore, the activation energies of grain ion conductivity were higher than for the composites sintered at higher temperatures and the conductivity of the grain boundary remains relatively low (see Table 2). At 900°C, the chemical modifications of the grain and boundary were completed. However, the mechanical stress inside the very large grains caused microcracks, which we observed on the SEM images. As a result, although the activation energies of the grain conductivity for the materials sintered at 800°C and 900°C were identical, the total conductivities after sintering at 900°C were lower than those of materials sintered at 800°C. Considering this result in the frame of the BLM, in particular referring to the equation (3) and inequality $\alpha\sigma_{gr}/\sigma_{gb} \ll 1$, and taking into account evidences of the SEM investigations indicating $\alpha(800°C) < \alpha(900°C)$, one can conclude that $\sigma_{gb}(800°C) > \sigma_{gb}(900°C)$. So, sintering at temperatures 900°C, in fact, causes the real ionic conductivity of grain boundary to be decreased compared to that of the materials sintered at 800°C. Various values for the activation energy of the total or grain ion conductivity have been reported in the literature [7, 8, 11, 13-15, 18-20, 22, 23, 25]. In the light of our experimental evidences, we think that the discrepancies in the values of activation energy determined and reported by various researchers resulted from differences in grain size and grain boundary due to the use of distinct processes to prepare them. We believe that different preparation protocol, especially long-time ball-milling, higher pressure used for formation of the pellets or prolonged sintering at higher temperatures might be the reason of various values of the ionic conductivity among many research papers. Further enhancement of electric properties is an important issue and it still under investigation to obtain materials characterized by the ionic conductivity exceeding $10^{-3}$ S·cm$^{-1}$.

## 4. Conclusions

The work presents the characterization of the ceramics formed in the system of $Li_{1.3}Al_{0.3}Ti_{1.7}(PO_4)_3$-$x$LiF with $0 \leq x \leq 0.3$. The total ion conductivity of the as-prepared polycrystalline $Li_{1.3}Al_{0.3}Ti_{1.7}(PO_4)_3$ is low due to presence of berlinite and amorphous phases containing aluminum with low electrical conductivities. Sintering at high temperature (700-900°C) modifies the phase composition of the grain boundary and decomposes the amorphous phase but not the berlinite. Therefore, the $Li_{1.3}Al_{0.3}Ti_{1.7}(PO_4)_3$ ceramics still exhibit poor total ion conductivity reaching a value $4.7 \times 10^{-5}$ S·cm$^{-1}$ at 30°C. Sintering of the $Li_{1.3}Al_{0.3}Ti_{1.7}(PO_4)_3$ with LiF additive causes three processes, each of which enhances the total ion conductivity: the growth of grains, the decomposition of amorphous or strongly disordered and AlPO$_4$ phases in grain boundary, accompanied by the formation of some lithium conducting phosphates. The best conducting material, LATP-0.1LiF sintered at 800°C, exhibits $\sigma_{tot} = 1.1 \times 10^{-4}$ S·cm$^{-1}$ at 30°C. The studied materials are interesting as a potential candidates for application as solid electrolytes in lithium ion battery technology.

**Acknowledgments**

This project has received funding from the European Union's Horizon 2020 research and innovation program under grant agreement No 731019 (EUSMI).